\pdfoutput=1
\RequirePackage{fix-cm}
\documentclass{svjour3}                     
\usepackage{graphicx}
\usepackage{amssymb}
\journalname{submitted}
\begin{document}

\title{Microwave properties of Nb/PdNi/Nb trilayers.}

\subtitle{Observation of flux flow in excess of Bardeen-Stephen theory.}


\author{	K. Torokhtii	\and
		N. Pompeo	\and
		C. Meneghini	\and
		C. Attanasio 	\and
		C. Cirillo		\and
		E. A. Ilyina	\and
		S. Sarti		\and
		E. Silva
}


\institute{	E. Silva (\email{silva@fis.uniroma3.it})	\and	C. Meneghini	\and	N. Pompeo	 \and	K. Torokhtii
			         	\at
              				Dipartimento di Fisica ``E. Amaldi" and CNISM, Universit\`a Roma Tre, Via della Vasca Navale 84, 00146 Roma, Italy\\
					\and
		C. Attanasio 	\and		C. Cirillo	\and		E. A. Ilyina
					\at
					CNR-SPIN and Dipartimento di Fisica ``E. Caianiello", Universit\`a degli Studi di Salerno, 84084 Fisciano (SA), Italy\\
					\and
		S. Sarti		\at
              				Dipartimento di Fisica, Universit\`a ``La Sapienza", 00185 Roma, Italy\\
}

\date{Oct. 9th, 2012}

\maketitle

\begin{abstract}
We combine wideband (1-20 GHz) Corbino disk and dielectric resonator (8.2 GHz) techniques to study the microwave properties in Nb/PdNi/Nb trilayers, grown by UHV dc magnetron sputtering, composed by Nb layers of nominal thickness $d_S$=15 nm, and a ferromagnetic PdNi layer of thickness $d_F$= 1, 2, 8 and 9 nm. We focus on the vortex state. Magnetic fields up to $H_{c2}$ were applied. The microwave resistivity at fixed $H/H_{c2}$ increases with $d_F$, eventually exceeding the Bardeen Stephen flux flow value.

\end{abstract}

\section{Introduction}
\label{intro}
Hybrid superconductor/ferromagnet structures have been the subject of intense attention in recent years \cite{buzdinRMP05,lyuksyutovAdP05}, due to their remarkable properties. Among the many properties studied, the simple flux flow state has been less under focus, even if its direct connection with the physics of the vortex core could led to uncovered results. One possible reason is that to reach the flow state in dc, large currents need to be injected in the samples, so that pinning phenomena, local heating and technical issues can make the experimental realization difficult. A feasible solution is to increase the frequency of an ac current up to frequencies large enough so that very small oscillations make the pinning profile irrelevant. In fact, many theories for the ac resistivity of superconductors in the vortex state \cite{GR,CC,brandtPRL91,MStheory} can be cast under the ``universal" expression \cite{pompeoPRB08}:
\begin{equation}
\label{eq:rhovm}
    \rho_{vm}=\rho_{vm,1}+\mathrm{i}\rho_{vm,2}=\rho_{\it ff}\frac{\varepsilon+\mathrm{i}\left(\nu/\bar{\nu}\right)}{1+\mathrm{i}\left(\nu/\bar{\nu}\right)}
\end{equation}
\noindent where $\bar{\nu}$ is a characteristic frequency and the $0\leq\varepsilon\leq 1$ is a measure of thermal activation phenomena. When $\nu\gg\bar{\nu}$, $\rho_{vm}\simeq \rho_{\it ff}$, irrespective of pinning and creep (whose relevance are within $\bar{\nu}$ and $\varepsilon$) and the flux-flow  (viscous drag only) resistivity is often described in terms of the Bardeen-Stephen (BS) model \cite{BS}:
\begin{equation}
\label{eq:rhoBS}
    \rho_{\it ff,BS}=c \rho_n {B}/{\mu_0 H_{c2}}
\end{equation}
where $\rho_n$ is the normal state resistivity, $B$ is the field induction, $H_{c2}$ is the temperature-dependent upper critical field and $c\lesssim 1$.

Aim of this paper is to exploit the microwave technique in order to verify to what extent Eq.(\ref{eq:rhoBS}) is satisfied in Nb/PdNi/Nb trilayers and in Nb. In Sec.\ref{exp} we present the sample structural characterization, the combined microwave measurements, and the results obtained. In Sec.\ref{disc} we discuss the results in the light of commonly accepted theories of flux flow, and we present experimental evidence for flux-flow resistivity in excess of the Bardeen-Stephen theory in S/F/S trilayers. A summary is presented in Sec.\ref{conc}.

\section{Experimental section}
\label{exp}
Measurements were performed on four Nb/Pd$_{0.84}$Ni$_{0.16}$/Nb trilayers. Samples have been grown on Si \cite{cirilloSUST11} and AlO$_2$ substrates by UHV dc diode magnetron sputtering. The thickness of each layer was measured by means of a quartz crystal monitor which, in turn, has been calibrated over a set of samples by low-angle x-ray reflectivity. Only the samples on Al$_2$O$_3$ have been considered for microwave measurements, to avoid complications due to the semiconducting substrate \cite{pompeoSUST05}. The nominal thickness of both Nb layers was always $d_{Nb}=$15 nm. The thickness of the ferromagnetic PdNi layers varied from 1 to 9 nm. A pure Nb sample of $d_{Nb}=$30 nm has been grown for comparison (labelled with $d_F=0$ in Table \ref{tabella}).
The film morphology of selected samples has been investigated by high resolution TEM (HRTEM) showing the film thickness close to the nominal value, excellent crystallinity of Nb layers and significant disorder in the PdNi layer suggesting interdiffusion of Nb into the PdNi layer.
The extended x-ray absorption spectroscopy (EXAFS) at the Nb K-edge is used to probe the local atomic structure on 30 nm Nb and selected Nb/PdNi/Nb trilayer samples (aged on time scales of 6 months). The EXAFS data analysis revealed a larger local disorder around Nb sites in trilayers with respect to pure Nb film. Preliminary EXAFS and HRTEM results coherently suggest some interdiffusion of top layer Nb into the middle PdNi layer. However some ageing effect may partially hinder HRTEM and EXAFS information, therefore further investigations are in progress and will be presented in a future publication.

All the microwave measurements here reported refer to a homogeneous set of samples, grown within a few days and measured in a comparable short time.

\begin{figure}[h]
\centerline{\includegraphics[height=6cm]{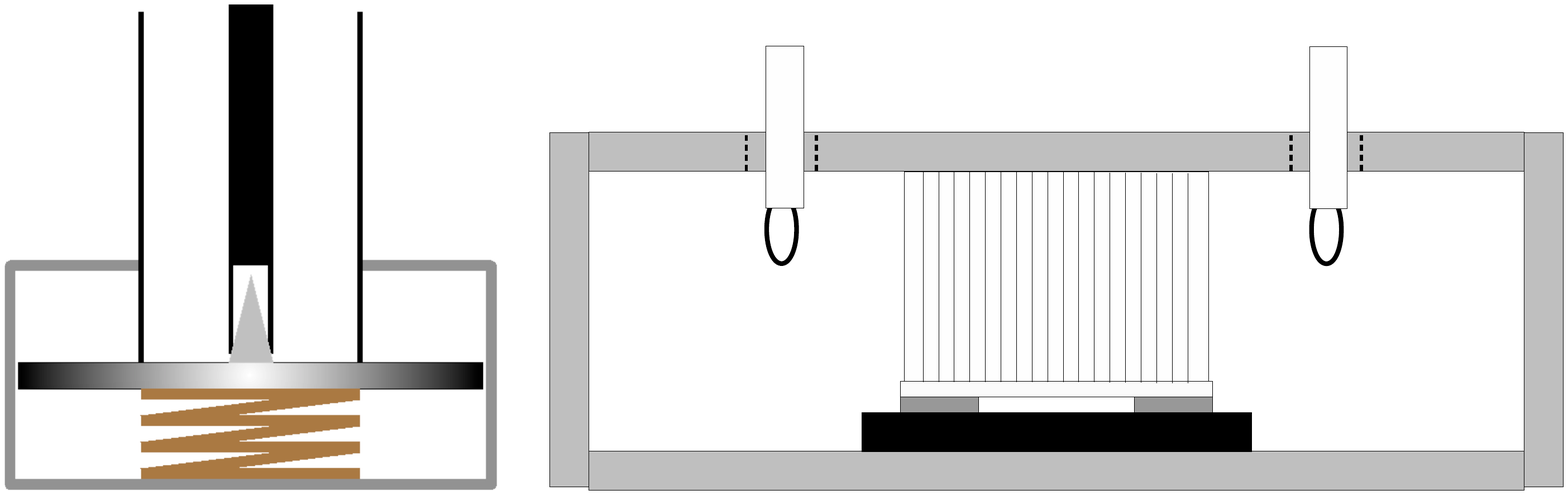}}
\vspace{-30mm}
  \caption{Left: sketch of the Corbino Disk setup; sample: shaded rectangle. Right: sketch of the Dielectric Resonator setup; sample: black rectangle.}
\label{figsetup}
\end{figure}

Table \ref{tabella} reports the critical temperature $T_c$ for our samples.
\begin{table}[h]
\centerline{
\begin{tabular}{l c c c c}
\hline
$d_F$ (nm) & $T_c$ (K) & Setup used & Measuring\\
  &   &   & temperature (K) \\
\hline
0 & 7.5 & CD  & 6.5\\
1 & 6.2 & CD & 5.3\\
2 & 5.1 & CD, DR & 4.4 \\
8 & 4.1 & CD, DR & 3.6\\
9 & 4.2 & DR & 3.6 \\
\hline
\end{tabular}
}
\caption{Main physical parameter of the samples investigated and information on the measurements here presented.}
\label{tabella}
\end{table}
We used a wideband (1-20 GHz) Corbino Disk (CD) setup \cite{silvaSUST11} to obtain the real part of the normalized microwave resistivity as a function of the frequency, $\rho_{1}(\nu)/\rho_n$ at selected fields, and a Dielectric Resonator (DR) technique \cite{torokhtiiPhCsub} to obtain $\rho_1(H)/\rho_n$ at the operating frequency $\nu_0\approx \;$ 8.2 GHz. The setups are sketched in Fig.\ref{figsetup}. A magnetic field up to $H_{c2}$ could be applied, perpendicular to the film plane. The temperature could be varied with $T\geq\;$3.3 K. In this paper we focus on homogeneous measurements taken in all samples at the same reduced temperature $t=T/T_c$ and reduced field $h=H/H_{c2}$.

\begin{figure}[h]
\centerline{\includegraphics[height=6cm]{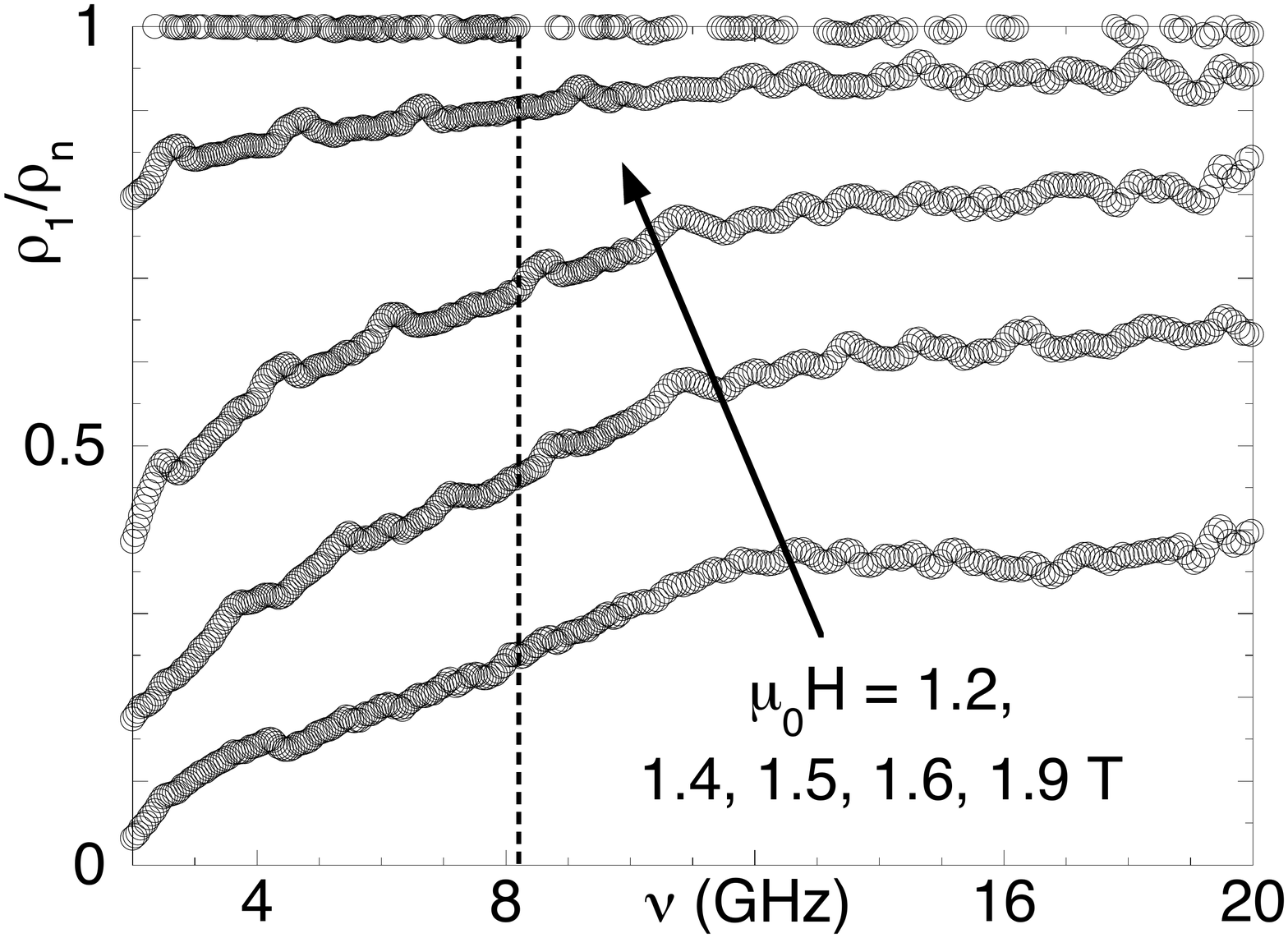}}
  \caption{Typical measurements of $\rho_1(\nu)/\rho_n$ taken with the CD. Here $H_c2=$1.9 T has been determined from the vanishing of $1-\rho_1/\rho_n$. The vertical dashed line is the cut at 8.2 GHz: the corresponding data points are compared in Fig.\ref{figCDDR} to the measurements obtained with the dielectric resonator.}
\label{figCD}
\end{figure}

The CD is particularly sensitive to various sources of perturbation, being a nonresonant technique. In particular, it is sensitive to mismatch in the coaxial line due to, e.g., temperature gradient. Such perturbations become less and less important with increasing dissipation levels, so that in the flux-flow regime of interest here this is not a large source of uncertainty. The fundamental advantage of the CD resides in the possibility to gain access to the full frequency dependence of the microwave resistivity. In Figure \ref{figCD} we report a sample measurement of $\rho_1(\nu)$ as measured in the Nb film at the lowest temperature attainable and selected fields. The typical increase of $\rho_1$ with $\nu$, consistent with Eq.(\ref{eq:rhovm}), is visible. The small peaks on top of the curves are manifestations of bistable behavior described previously \cite{pompeoPhC10}. With increasing frequency, $\rho_1$ reaches a plateau. Consistently with Eq.(\ref{eq:rhovm}), the plateau is a measure of the true flux-flow resistivity. Similarly, the frequency at the midpoint of $\rho_1(\nu\ll\bar{\nu})-\rho_
1(\nu=0)$ is a measure of the characteristic vortex frequency $\bar{\nu}$. Thus, directly from the data one can estimate both the effect of pinning (from $\bar{\nu}$) and the value of the flux flow resistivity.

The DR presents advantages and drawbacks completely complementary to the CD. It is a highly-sensitive technique, albeit limited to a single frequency (here 8.2 GHz, the resonant frequency of the resonator) and to dissipation level low enough in order to avoid excessive degradation of the quality factor $Q$. The DR used in the measurements here presented \cite{torokhtiiPhCsub} is based upon a rutile (TiO$_2$) cylindrical puck shielded by a OFHC Cu enclosure. The film was placed on a base of the dielectric cylinder. Measurements of the field variation of $Q$ yielded the field increase of $\rho_1(H)$. While the field variations of the quality factor were rather robust with respect to thermal stability and thermal gradients, the same did not hold for the resonant frequency, which is sensitive even to small changes in the He flux. Thus, simultaneous measurements of the shift of the resonant frequency required extra long thermal stabilization, and were performed only on a subset of the measurements.

\begin{figure}[h]
\centerline{\includegraphics[height=6cm]{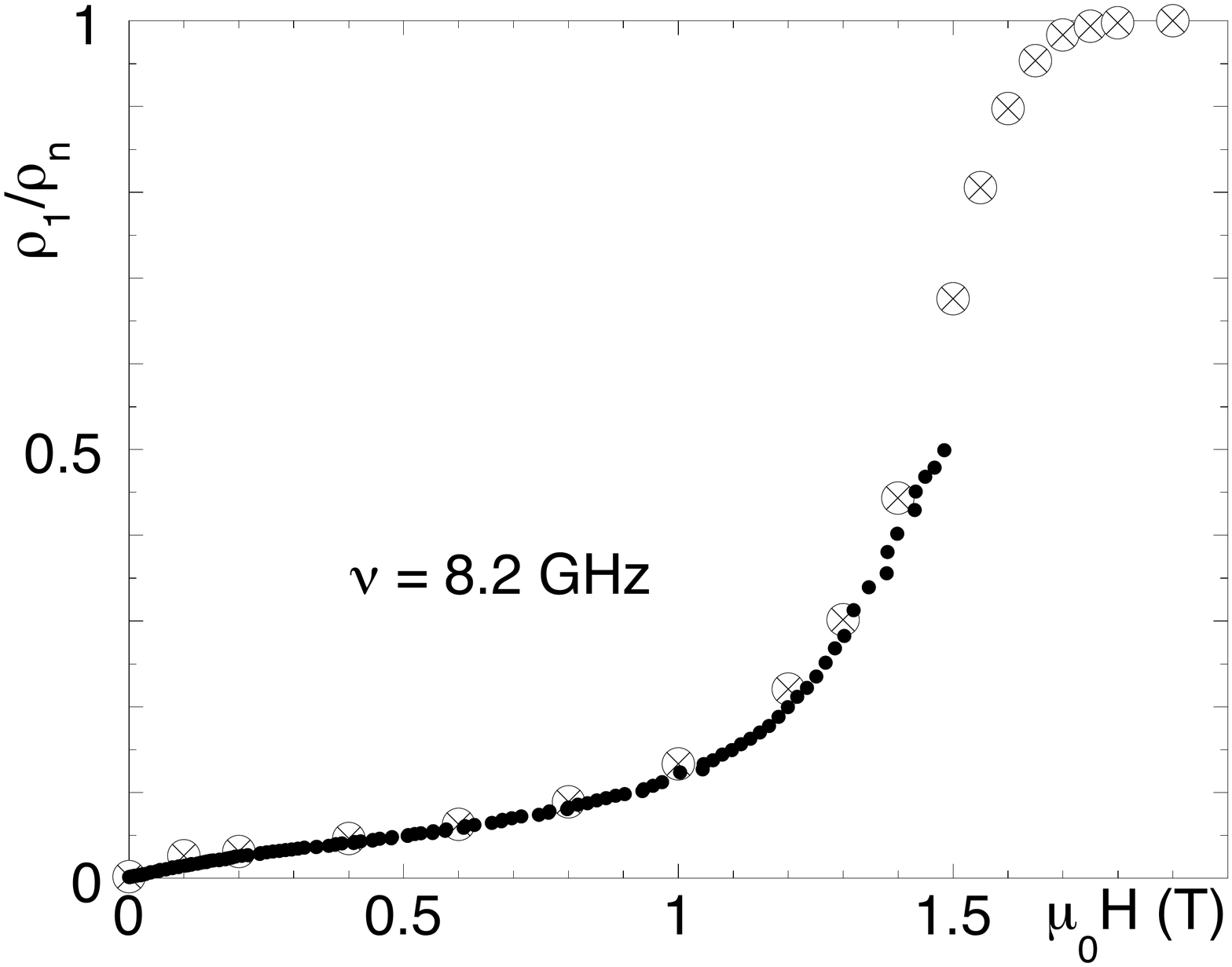}}
  \caption{Typical measurements of $\rho_1(H)/\rho_n$ taken with the DR; crossed circles are measurements taken with the CD at 8.2 GHz (from Fig.\ref{figCD}).}
\label{figCDDR}
\end{figure}

A comparison of the measurements taken with the DR and the CD is reported in Figure \ref{figCDDR}. There, we report $\rho_1(H)$ at 8.2 GHz in the Nb sample, as measured with the DR and as obtained with the CD (dashed line in Figure \ref{figCD}).

In the following we discuss measurements of the real part of the resistivity $\rho_1$, taken at $t\simeq 0.86$ and  $h\simeq 0.5$, that is in a dissipative regime where both the CD and the DR have sufficient sensitivity.
%
%
%

\section{Discussion}
\label{disc}
Since focus of the present results is the compatibility of $\rho_{\it ff}$ with existing models, we note preliminarily that the measured $\rho_1$ is in all cases an underestimate of $\rho_{\it ff}$: $\rho_1\leq\rho_{\it ff}$. In order to keep as much contact as possible to the raw data, it will prove sufficient to discuss directly $\rho_1$, without attempting further data transformation.

\begin{figure}[h]
\centerline{\includegraphics[height=6.cm]{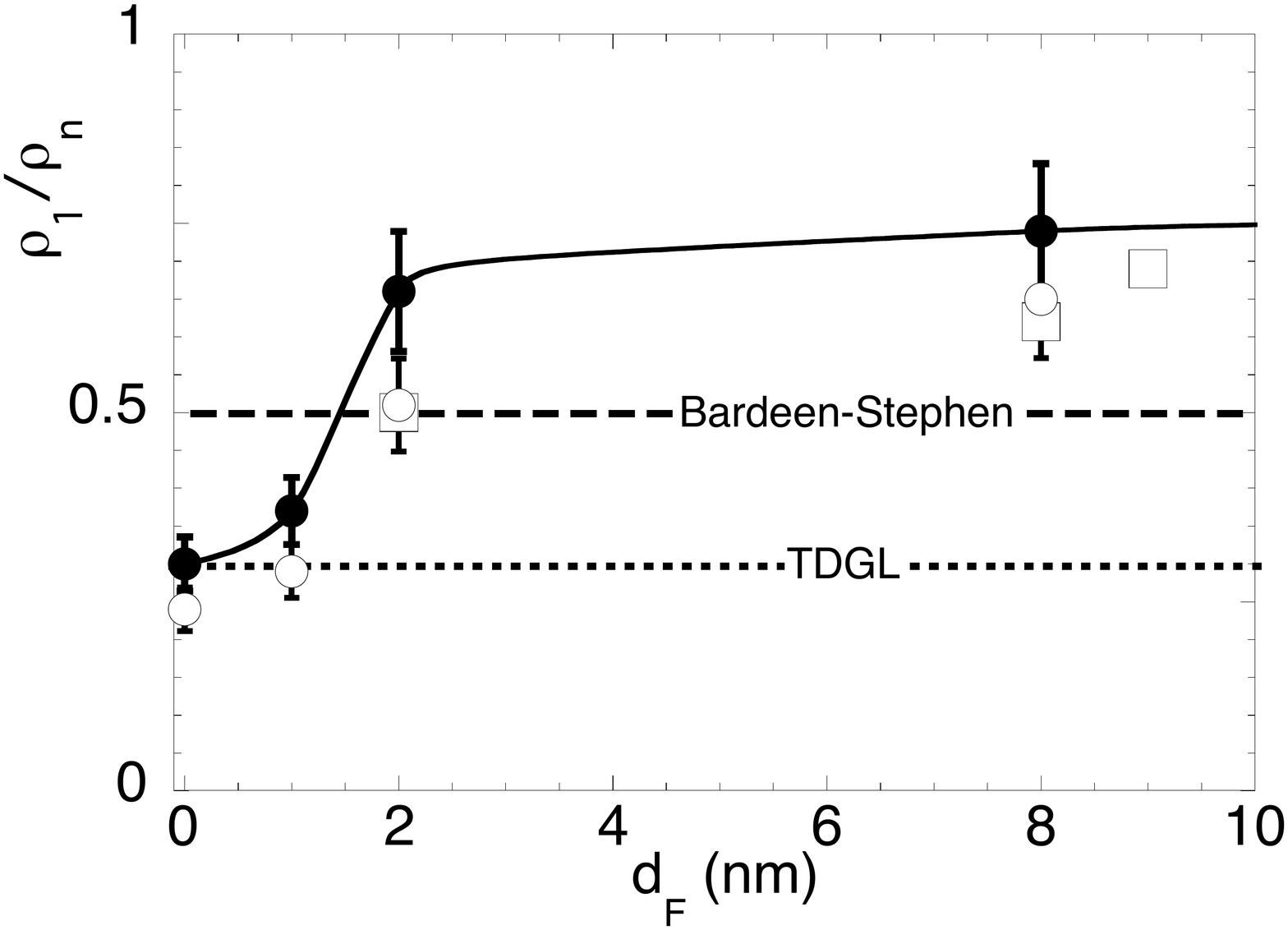}}
\vspace{-3mm}
  \caption{Measurements of $\rho_{1}/\rho_n$ at $t=T/T_c\simeq$0.86 and $h=H/H_{c2}\simeq$0.5 as a function of $d_F$. Full dots: ``plateau" resistivity (approximation to $\rho_{\it ff}$) as measured with the Corbino Disk. The Bardeen-Stephen limit is clearly not obeyed, and eventually exceeded as $d_F$ grows. The TDGL calculation \cite{troyPRB93,liangPRB2010} applies only at low $d_F$. Continuous line is a guide to the eye. Open circles: $\rho_1/\rho_n$ at 8.2 GHz from the Corbino Disk measurements at 8.2 GHz. Open squares: $\rho_1/\rho_n$ obtained with the Dielectric Resonator at 8.2 GHz. }
\label{figrhoff}
\end{figure}

From the measurements taken with the DR one directly obtains $\rho_1/\rho_n$ at 8.2 GHz. From the measurements with the CD, we selected $\rho_1$ at 8.2 GHz (for comparison with the DR), and in the plateau region of $\rho_1(\nu)$, in order to approximate $\rho_{\it ff}$, see Eq.(\ref{eq:rhovm}). In Figure \ref{figrhoff} we report the data for $\rho_1/\rho_n$ at $h\simeq 0.5$ and $t\simeq 0.86$ as a function of $d_F$. We first notice that at large $d_F$, $\rho_1(8\;$GHz$)/\rho_n$ approaches $\rho_1/\rho_n$ at the plateau. This is a direct indication that vortex pinning becomes less effective as the ferromagnetic layers thickness increases. Second, measurements with CD and DR at the same frequency are in good agreement, as already noticed. Third, $\rho_1/\rho_n$ clearly increase with $d_F$, and reach values well above the BS limit at large $d_F$. This is unexpected, since the BS value should be an upper limit (and more so at high $h$, since $\rho_{\it ff}\rightarrow \rho_n H/ H_{c2}$).

The horizontal dashed line is the BS prediction, Eq.(\ref{eq:rhoBS}). The horizontal dotted line is a full calculation within a time-dependent Ginzburg Landau (TDGL) framework \cite{troyPRB93} as reported in \cite{liangPRB2010}, where the dc flux flow resistivity was found to follow closely the TDGL calculation. As it can be seen, at low $d_F$ (including pure Nb) the TDGL model is compatible with the data. At large $d_F$, the experimental $\rho_1$ exceeds the BS value. Since the true $\rho_{\it ff}$ is not smaller than the measured $\rho_1$, this is a direct demonstration that the flux flow resistivity in our superconductor/ferromagnet/superconductor trilayers exceeds the ratio $\rho_n H/H_{c2}$. This observation remains unexplained, and points to new phenomena in the vortex cores.

\section{Conclusions}
\label{conc}

In conclusion, we have shown that microwave measurements allow to access the flux flow resistivity at subcritical currents. We have shown that $\rho_{\it ff}$ never obeys the BS model in a regime where the model itself is supposed to hold. At small $d_F$, and in the pure Nb sample, $\rho_{\it ff}<\rho_{\it ff,BS}$. This observation is compatible with existing models for the flux flow resistivity based upon microscopic \cite{LO} or TDGL \cite{troyPRB93} calculations, and they are in agreement with dc data \cite{liangPRB2010}. By contrast, at relatively large ferromagnetic thickness one finds $\rho_{\it ff}>\rho_{\it ff,BS}$ at $H/H_{c2}=0.5$. This unexpected behavior is not explained, to our knowledge, by existing models for $\rho_{\it ff}$. Further investigations are in progress to unveil the field and temperature dependence of $\rho_{\it ff}$ in superconductor/ferromagnet hybrids.\\

\begin{acknowledgements}
This work has been partially supported by an Italian MIUR-PRIN 2007 project.
\end{acknowledgements}

\end{document}